\documentclass[prb,twocolumn,showpacs,superscriptaddress,amsmath]{revtex4-1}
\usepackage{amssymb}
\usepackage{graphicx}
\usepackage{bm}
\usepackage{color}

\begin{document}
\title{Bulk and surface electronic properties of SmB$_6$: a hard x-ray
photoelectron spectroscopy study}

\author{Y. Utsumi}
\email{yuki.utsumi@synchrotron-soleil.fr}
\affiliation{Max Planck Institute for Chemical Physics of Solids,
N\"othnitzer Stra\ss e 40, 01187 Dresden, Germany}
\author{D. Kasinathan}
\affiliation{Max Planck Institute for Chemical Physics of Solids,
N\"othnitzer Stra\ss e 40, 01187 Dresden, Germany}
\author{K-T. Ko}
\affiliation{Max Planck Institute for Chemical Physics of Solids,
N\"othnitzer Stra\ss e 40, 01187 Dresden, Germany}
\author{S. Agrestini}
\affiliation{Max Planck Institute for Chemical Physics of Solids,
N\"othnitzer Stra\ss e 40, 01187 Dresden, Germany}
\author{M. W. Haverkort}
\altaffiliation[Present Address: ]{Institute for theoretical physics, Heidelberg University, 69120 Heidelberg, Germany}
\affiliation{Max Planck Institute for Chemical Physics of Solids,
N\"othnitzer Stra\ss e 40, 01187 Dresden, Germany}
\author{S. Wirth}
\affiliation{Max Planck Institute for Chemical Physics of Solids,
N\"othnitzer Stra\ss e 40, 01187 Dresden, Germany}
\author{Y-H. Wu}
\affiliation{National Synchrotron Radiation Research Center, 101 Hsin-Ann Road, Hsinchu 30077, Taiwan}
\author{K-D. Tsuei}
\affiliation{National Synchrotron Radiation Research Center, 101 Hsin-Ann Road, Hsinchu 30077, Taiwan}
\author{D-J. Kim}
\affiliation{Department of Physics and Astronomy, University of California, Irvine, CA 92697, USA}
\author{Z. Fisk}
\affiliation{Department of Physics and Astronomy, University of California, Irvine, CA 92697, USA}
\author{A. Tanaka}
\affiliation{Department of Quantum Matter, AdSM, Hiroshima University, Higashi-Hiroshima 739-8530, Japan}
\author{P. Thalmeier}
\affiliation{Max Planck Institute for Chemical Physics of Solids,
N\"othnitzer Stra\ss e 40, 01187 Dresden, Germany}
\author{L. H. Tjeng}
\affiliation{Max Planck Institute for Chemical Physics of Solids,
N\"othnitzer Stra\ss e 40, 01187 Dresden, Germany}

\date{\today}

\begin{abstract}
We have carried out bulk-sensitive hard x-ray photoelectron spectroscopy (HAXPES) measurements on \textit{in-situ} 
cleaved and \textit{ex-situ} polished SmB$_6$ single crystals. Using the multiplet-structure in the Sm $3d$ core level 
spectra, we determined reliably that the valence of Sm in bulk SmB$_6$ is close to 2.55 at $\sim$5 K. 
Temperature dependent measurements revealed that the Sm valence gradually increases to 2.64 
at 300 K. From a detailed line shape analysis we can clearly observe that not only the $J=0$ 
but also the $J=1$ state of the Sm $4f^6$ configuration becomes occupied at elevated temperatures.  
Making use of the polarization dependence, we were able to identify and extract the 
Sm $4f$ spectral weight of the bulk material. Finally, we revealed that the oxidized or chemically damaged surface region of the \textit{ex-situ} polished SmB$_6$ single crystal is surprisingly thin, about 1 nm only.

\end{abstract}


\maketitle

\section{Introduction}

The interplay of strong spin-orbit coupling and electron-electron correlations in rare earth compounds has recently 
been shown theoretically to allow for the emergence of topologically nontrivial surface bands, thereby merging 
the fields of strongly correlated systems and Kondo physics with topology. A minimum model consisting of localized 
$f$-electrons and dispersive conduction electrons with opposite parity provides us a topological $f$-electron system 
that hosts topologically protected metallic surface states within a hybridization gap, i.e. a topological Kondo insulator 
\cite{Dzero2010}.

In this context, it was proposed \cite{Dzero2010,Takimoto2011,Dzero2012,Lu2013,Dzero2013,Alexandrov2013} 
that the Kondo insulator, or intermediate valent system, SmB$_6$ is a good candidate material to qualify as the first strongly 
correlated topological insulator. Indeed, the robust metallicity which is attributed to a topologically protected surface 
state could be a promising explanation for the long-standing mysterious low-temperature residual conductivity of 
SmB$_6$ \cite{Allen1979,Gorshunov1999,Flachbart2001}. SmB$_6$ has therefore triggered a tremendous 
renaissance in recent years, and many research efforts have been made to establish the topological nature of the 
material using a wide range of experimental methods, e.g. angle-resolved photoelectron spectroscopy (ARPES) 
\cite{Xu2013,Zhu2013,Neupane2013,Jiang2013,Denlinger2014,Xu2014}, scanning tunneling spectroscopy 
\cite{Yee2013,Roessler2014,Ruan2014,Roessler2016, Jiao2016}, resistivity and surface conductance measurements 
\cite{Hatnean2013, Zhang2013,Kim2013,Wolgast2013,Kim2014,Wolgast2015,Thomas2016,Nakajima2016}, 
and high pressure experiments \cite{Butch2016,Sun2016,Zhou2016}. A recent special issue with 
foreword provides an excellent overview of the field \cite{Allen2016}.

SmB$_6$ is an intermediate valent compound where the valence number ($v$) of Sm ion varies between 2+ and 3+ as 
first observed by x-ray absorption experiments \cite{Vain65}. An early magnetic susceptibility study \cite{Cohen1970} 
hinted at a valence of $v\sim2.6$ while a subsequent x-ray photoelectron spectroscopy (XPS) experiment  
\cite{Allen1980} extracted $v\sim 2.7$ at room temperature. Using Sm $L_3$ x-ray absorption spectroscopy, 
the valence numbers $v =$ 2.6-2.65 \cite{Vain65}, 2.53 at $T=$ 4.2 K \cite{Tarascon1980}, 
and 2.52 at $T=$ 2 K \cite{Mizumaki2009} were determined. A Sm $L_{\gamma 4}$ emission spectroscopy 
study found $v =$ 2.65 at room temperature \cite{Hayashi2013}, and a very recent take-off angle 
photoemission study yielded $v =$ 2.48 at 150 K for the bulk \cite{Lutz2016}. 
The Sm valence is an important issue for the theory of the proposed topological character of SmB$_6$. 
While an \textit{ab-initio} based study including the full $4f$-orbital basis predicts the topological insulator phase 
with $v \approx$ 2.5 \cite{Lu2013}, model calculations for materials with cubic symmetry including only the 
$\Gamma_8$ quartet states proposed a phase diagram in which SmB$_6$ is expected 
\cite{Dzero2013,Alexandrov2013} to be a band insulator for $v <$ 2.56,  and a topological Kondo insulator 
when 2.56 $< v <$ 3. 

\begin{figure*}[t]
\includegraphics[scale=0.6]{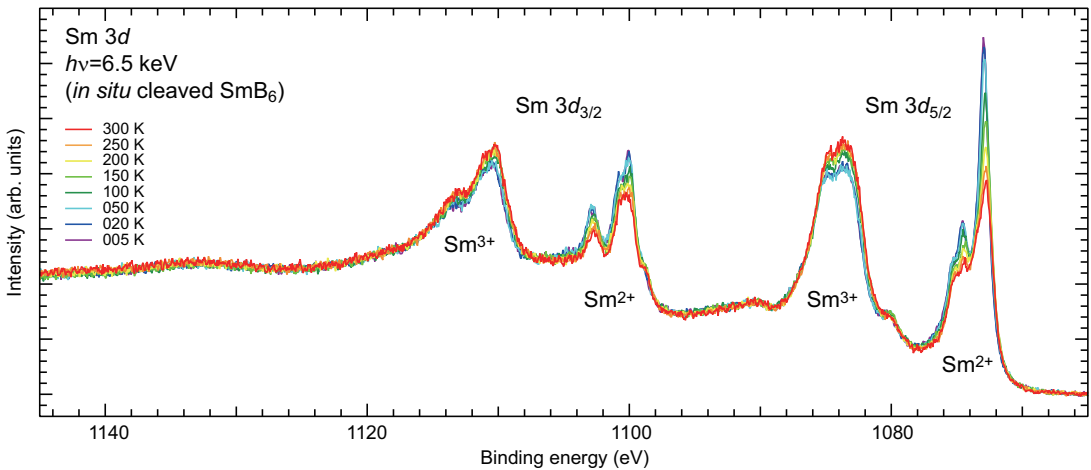}
\caption{(Color online) Temperature dependence of the Sm 3$d$ spectra of an \textit{in-situ} cleaved 
SmB$_6$ single crystal. With increasing temperature, the intensity of the Sm$^{2+}$ (Sm$^{3+}$) 
component decreases (increases).} \label{fig.1}
\end{figure*}

As mentioned above, several experiments have been performed on SmB$_6$ to estimate the Sm valence. 
However, the obtained value varies depending on the experimental methods. Here, we performed bulk sensitive 
hard x-ray photoelectron spectroscopy (HAXPES) to collect the Sm $3d$ core level spectra from which the Sm 
valence can be determined \cite{Yamasaki2005,Yamasaki2007, Lutz2016}. We utilized the intensities of the 
multiplet structure of the Sm$^{2+}$ and Sm$^{3+}$ features, and by doing so, we did not need to model 
the background, and were therefore able to extract more reliably the ratio between the Sm$^{2+}$ and 
Sm$^{3+}$ signals. Since many of the reported resistivity and surface conductance experiments on SmB$_6$ 
\cite{Hatnean2013,Zhang2013,Kim2013,Wolgast2013,Kim2014,Wolgast2015, Thomas2016,Nakajima2016,Zhou2016} 
have been carried out at ambient conditions or on samples which were prepared at such conditions, there is 
also a need to evaluate the effect of ambient conditions on the SmB$_6$ surface. We therefore performed 
HAXPES on \textit{in-situ} cleaved SmB$_6$ and \textit{ex-situ} polished SmB$_6$ and compared the results.

\section{Experimental}
\label{sec-exp}
The experiments have been carried out at the Max-Planck-NSRRC HAXPES station at 
the Taiwan undulator beamline BL12XU at SPring-8, Japan. The photon beam with $h\nu \sim6.5$ keV is 
linearly polarized with the electrical field vector in the plane of the storage ring (i.e. horizontal). Two MB 
Scientific A-1 HE hemispherical analyzers have been used in two different geometries: The first analyzer 
was mounted horizontally and parallel to the electrical field vector of the photon beam. The second analyzer 
was in the vertical geometry, perpendicular to the electrical field vector and the Poynting vector of the beam. 
A detailed description of the experimental setup can be found in Ref.\ \onlinecite{Weinen2015}. The overall 
energy resolution was set to $\sim$170 meV and the zero of the binding energy of the photoelectrons was 
determined using the Fermi edge of a gold film. The SmB$_6$ single crystals used in our study were grown 
by the aluminium flux method \cite{Kim2014}. One single crystal was cleaved \textit{in-situ} under ultrahigh 
vacuum conditions (better than $3\times10^{-10}$ mbar). A second single crystal was mirror-polished with 
an Al$_2$O$_3$ polishing pad, cleaned using diluted HCl for 2 minutes, rinsed with isopropanol, and 
subsequently transferred into the ultrahigh vacuum system. A detailed description of the polishing 
procedure can be found in Ref.\ \onlinecite{Kim2014}.

\section{Results and Discussions}
\subsection{Sm valence}

Figure \ref{fig.1} shows the Sm 3$d$ core level spectra of \textit{in-situ} cleaved SmB$_6$ for temperatures 
ranging from 5 K to 300 K. The Sm 3$d$ spectra are split into a 3$d_{5/2}$ and a 3$d_{3/2}$ branch due 
to the spin-orbit interaction. Each of these branches is further split into the so-called Sm$^{2+}$ (4$f^6$) 
and Sm$^{3+}$ (4$f^5$) components which represent the Sm $4f^6 \rightarrow \underline{c}~4f^6 + e$ 
and the Sm $4f^5 \rightarrow \underline{c}~4f^5 + e$ transitions, respectively, where $\underline{c}$ 
denotes a $3d$ core hole and $e$ the outgoing photoelectron. With increasing temperature, the intensity 
of the Sm$^{2+}$ (Sm$^{3+}$) component gradually decreases (increases) and consequently, the 
mean-valence $v$ of Sm moves towards becoming more trivalent. 
We would like to note that there were no detectable degradation effects of the
sample surface after the temperature cycle, see Appendix A.

\begin{figure*}[t]
\includegraphics[scale=0.6]{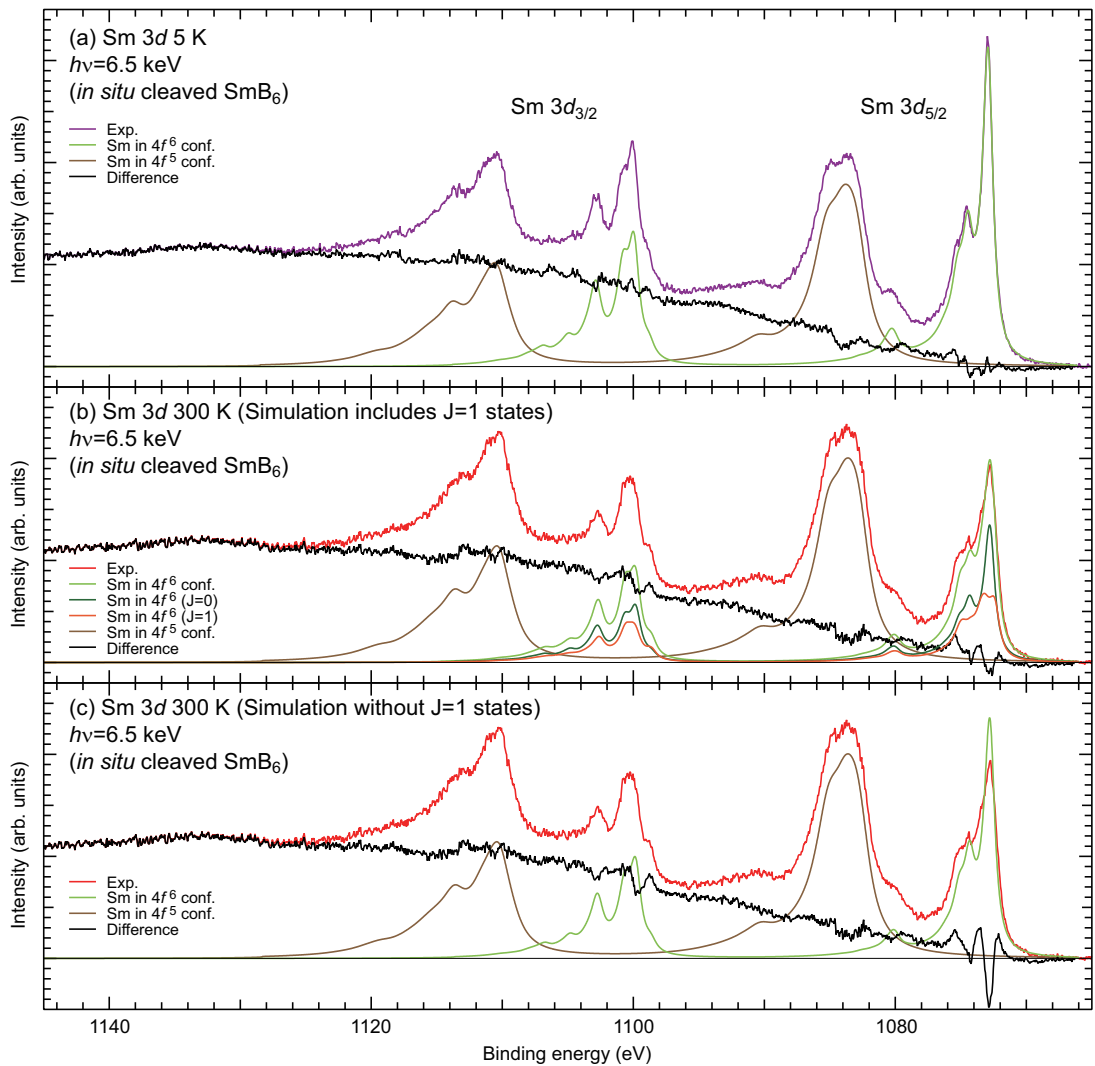}
\caption{(Color online) Multiplet analysis of the Sm 3$d$ spectra of the \textit{in-situ} cleaved 
SmB$_6$ sample. (a) $T=$ 5 K, (b) $T=$ 300 K including the Boltzmann occupation of the $J=1$ states of the 
Sm $^{2+}$ ($4f^6$) configuration in the simulation, (c) $T=$ 300 K without the $J=1$ states of the Sm $4f^6$. 
The experimental spectra at 5 K and at 300 K are presented by the purple and red lines, respectively. 
The simulations for the Sm$^{2+}$ and Sm$^{3+}$ components are displayed by the green and brown
lines, and a break down of the $J=0$ and $J=1$ components of the Sm $^{2+}$ ($4f^6$)
configuration by the dark green and orange lines, respectively. Black lines represent the inelastic background
signal and were obtained by subtracting the simulated multiplet structure from the experimental spectra.}
\label{fig.2}
\end{figure*}

In order to obtain $v$ quantitatively, a simulation analysis was performed on the spectra by carrying out atomic 
full-multiplet calculations to account for the lineshape of the Sm $3d$ core level spectra \cite{Haverkort2012, Tanaka1994}. 
Crystal field effects are not taken into account since the corresponding energy splittings are minute
compared to the lifetime broadening of the core-hole final states. The hybridization between the  Sm$^{2+}$ and 
Sm$^{3+}$ core hole final states is neglected in view of the fact that their energy separation is much larger than
the hopping integral between the $4f^6$ $J$=0 and $4f^5$ $J$=5/2 configurations which is very small due to both 
the contracted radial wavefunctions of the Sm 4$f$ and fractional-parentage matrix element effects 
\cite{Sawatzky2016}.
The calculated spectra are convoluted with a Lorentzian function for lifetime broadening and a Gaussian to 
account for the instrumental resolution. The experimental spectra at a given temperature are then fitted 
by adjusting the weights of the calculated Sm$^{2+}$ and Sm$^{3+}$ components such that in the 
difference spectrum between the experimental and calculated spectra the fingerprints of the Sm$^{2+}$ 
and Sm$^{3+}$ multiplet structures are minimized. The broadening parameters and as well as
the values used for the Coulomb and exchange multiplet interactions are listed in Ref. \onlinecite{Parameters}.

The results for $T=$ 5 K are shown in Fig.2(a). The experimental spectrum taken at $\sim$5 K (purple line) 
subtracted by the best fit for the Sm$^{2+}$ (green line) and Sm$^{3+}$ components 
(brown line) produces a difference spectrum (black line) which shows a gently sloping 
background plus some residual wiggling features which originate mostly from tiny 
deviations in the peak positions and peak widths of the multiplet structures.  A Sm mean-valence of $v =$ 2.55 is extracted from this spectrum by using formula 
$v$=2+$I_{3+}$/($I_{2+}$+$I_{3+}$). Here, $I_{2+}$ and $I_{3+}$ denote the integrated spectral 
intensities of the Sm$^{2+}$ and Sm$^{3+}$  simulated spectra, respectively, optimized to fit the experimental spectrum. 

In the simulations for the higher temperature spectra, we allow for the Boltzmann occupation
of the excited states of the Sm. Fig. 2(b) displays the results for the $T=$ 300 K spectrum. Here we can notice
that not only the $J=0$ (dark green line) but also the $J=1$ (orange line) state of the Sm$^{2+}$ ($4f^6$) 
configuration contributes to the spectrum. The energy splitting between the $J=0$ and $J=1$ states was set 
to 35 meV by fine tuning the $4f$ spin-orbit and multiplet interactions\cite{Parameters} as to match the 
energy splitting found from inelastic neutron scattering experiments\cite{Alekseev1993, Alekseev1995} resulting 
in about 57\% occupation for the $J=0$ and 43\% for the $J=1$ states at room temperature. The difference between 
the experimental spectrum and the multiplet calculation is a gently sloping background curve, similarly smooth 
like in the 5 K case, demonstrating the validity of the analysis procedure. We stress that in the simulation we 
cannot omit the $J=1$ Boltzmann occupation. This is clearly revealed by Fig. 2(c), which shows the poor match 
between the $J=0$ only simulation and the experimental spectrum for the Sm$^{2+}$ $3d_{5/2}$. The 
deviations can also be observed as strong wiggles in the difference spectrum between the experimental and the multiplet calculation. 

We would like to remark that for the Sm$^{3+}$ part of the spectrum, the simulations yield a temperature
independent line shape for the temperatures considered here. The energy splitting between $J$=5/2 and $J$=7/2 
multiplets is too large to cause an appreciable Boltzmann occupation of the higher lying $J$=7/2, so that the spectrum 
is given primarily by the lower lying $J$=5/2. Inclusion of a cubic crystal field will also not produce a temperature effect,
due to the fact that the $\Gamma_8$ and $\Gamma_7$ crystal field states originate from the same $J$ quantum 
number, while at the same time the crystal field energy scale is about two orders of magnitude 
smaller than that of the inverse life time of the $3d$ core-hole, i.e. any tiny spectral changes due to the crystal
field are washed out by the core-hole life time broadening, see Appendix B.

\begin{figure}[t]
\includegraphics[scale=0.6]{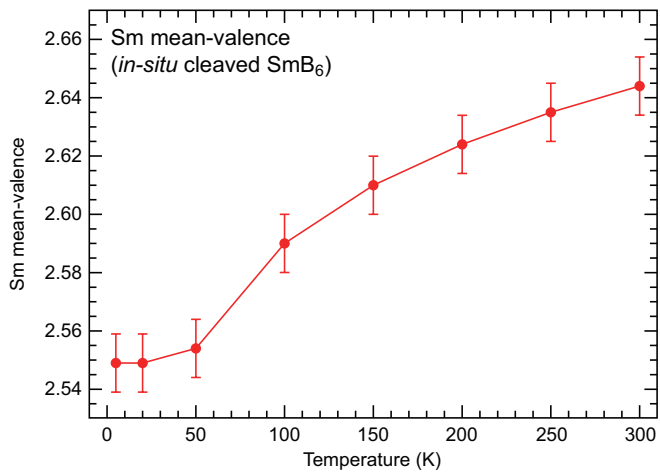}
\caption{(Color online) Temperature dependence of the Sm mean-valence of the \textit{in-situ} cleaved 
SmB$_6$ sample.}
\label{fig.3}
\end{figure}

Applying this procedure to spectra taken at other temperatures, allowed for a determination of the Sm 
mean-valence as a function of temperature. The results are plotted in the main panel of Fig.3, revealing 
a gradual increase of the Sm valence to a value of $v =$ 2.64 at 300 K.
In general, our findings for the Sm valence and its temperature dependence are consistent with the results 
reported in earlier Sm-$L_{2, 3}$ x-ray absorption spectroscopy (XAS) and Sm $L_{\gamma 4}$ emission 
spectroscopy studies \cite{Tarascon1980, Mizumaki2009, Hayashi2013}. However, our
experimental method and analysis are different with implications for the reliability of the extracted
values of the valence. The Sm$^{2+}$ and Sm$^{3+}$ components in our photoemission 
core level spectra are well separated, more so than in the Sm-$L_{2, 3}$ and $L_{\gamma 4}$ spectra. 
In addition, the presence of sharp multiplet structures in the Sm $3d$ spectra allows us to unambiguously 
assign the Sm$^{2+}$ and Sm$^{3+}$ components, such that their integrated intensities can be determined 
without having to model the background. In this way, we also ensure that the multiplet structures 
are fitted without violating the atomic 3$d_{5/2}$ and $3d_{3/2}$ branching ratio 
(see the Appendix C).
All this adds to the reliability of the valence determination by 
performing HAXPES on the $3d$ level. In comparing our HAXPES results with a recent HAXPES take-off 
angle study carried out at 150 K \cite{Lutz2016}, we would like to note that we have found quite a higher 
value for the valence, namely $v =$ 2.61 at 150 K, while the take-off angle HAXPES 
provided a value of only $v =$ 2.48. Perhaps this is related to the fact that the take-off angle HAXPES study 
has put more weight in getting a good simulation of the surface sensitive part of the data and thus less 
on the bulk properties.

One of the interesting findings here is that the low temperature valence of $v=$ 2.55 is very close to the 
border between SmB$_6$ being a band insulator (for $v <$ 2.56) or a topological Kondo insulator 
(for 2.56 $< v <$ 3) as pointed out in Refs. \onlinecite{Dzero2013,Alexandrov2013}. If we take these 
numbers seriously, then it is in fact not clear at all that SmB$_6$ can be expected to be a strongly correlated 
topological insulator. However,  the critical value $v_c=2.56$ that separates trivial and topological insulator 
depends on numerous model parameters and therefore may be subject to fine-tuning. Consequently 
further investigations, both theoretical and experimental, are clearly warranted.

Another important aspect is the increasing valence with temperature. This effect even outweighs thermal 
expansion, \textit{i.e.} the increasing presence of Sm$^{3+}$ (being smaller than  Sm$^{2+}$) causes 
the lattice constant to shrink with temperature (and correspondingly the linear thermal expansion coefficient 
to have negative values) for temperatures as high as 150 K \cite{Tarascon1980,Mandrus1994}. The valence 
is related to the number of $4f$ holes (in the degenerate $J=5/2$ state) by $n_f^h(T)=v(T)-2$. Without 
considering hybridization it becomes entropically favorable to occupy the more degenerate Sm$^{3+}$ 
$(J=5/2)$ hole states  instead of the Sm$^{2+}$ $(J=0)$ singlet state to decrease the free energy. 
Therefore $n_f^h(T)$ and hence the valence $v(T)$ increase with temperature. In a more microscopic 
picture including the hybridization, a part of the hole spectral weight is pushed above the Fermi level, 
which leads to a decrease in $n_f^h(T)$ when temperature decreases. This is due to the formation of 
the bound state of 4$f$ hole with a conduction electron as in the case \cite{Sakai1992} of Yb$^{3+}$.

\subsection{Valence Band}

\begin{figure}[t!]
\includegraphics[scale=0.6]{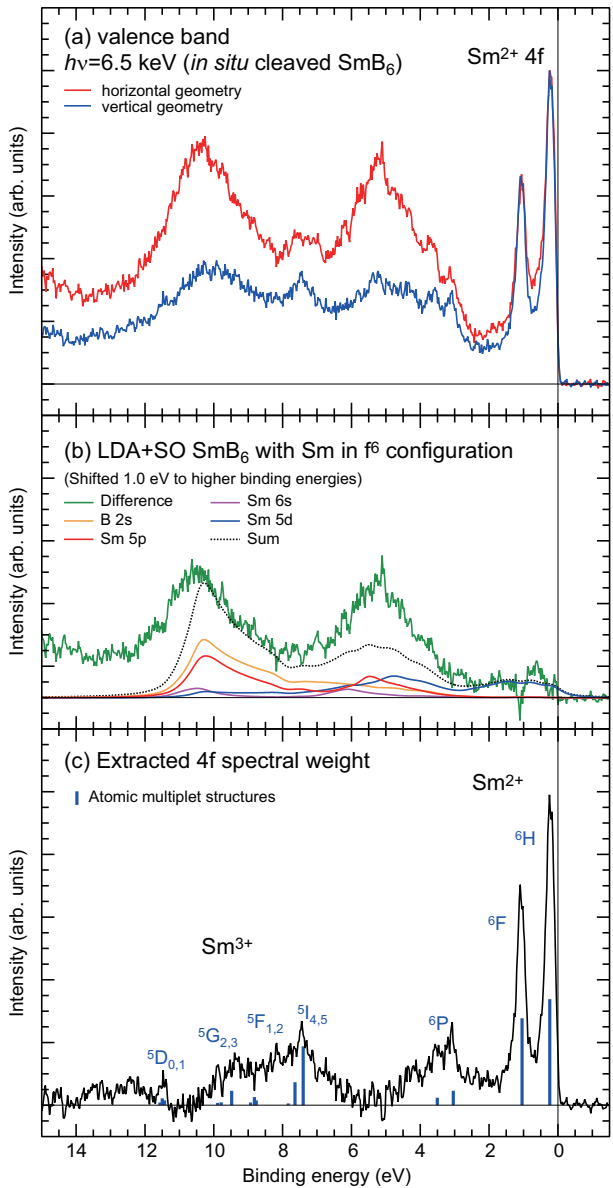}
\caption{(Color online) (a) Experimental valence band spectra of \textit{in-situ} cleaved SmB$_6$ 
measured in the horizontal (red) and vertical (blue) geometry at $T=$ 50 K. The spectra are normalized 
to the height of the Sm$^{2+}$ 4$f$ peaks. (b) Difference between the horizontal and vertical geometry
spectra together with the B $2s$ and Sm $5p$, $5d$, $6s$ partial density of states 
from a non-magnetic band structure calculation with Sm in the $f^6$ configuration. The densities of states 
are displayed with a shift of 1 eV towards higher binding energies and weighted with the 
photo-ionization cross-section factors as explained in the text. 
(c) The experimental valence band spectrum after a weighted (see text) subtraction of the difference 
spectrum (b) (black line), together with the assignment of the atomic multiplet structures (blue sticks and
labels) \cite{Denlinger2014}.} 
\label{fig.4}
\end{figure}

The large inelastic mean-free path of electrons with kinetic energies of several keV \cite{Tanuma1994} 
provides an opportunity to collect photoemission spectra that are representative of the bulk material by 
carrying out experiments using hard x-ray photons. The spectra from such HAXPES experiments can 
then be used as a reference in a comparison with spectra taken at lower photon energies in order to 
identify features that may originate from the surface region of the sample. In particular, the contribution 
of the surface may become significant if ultra-violet photon energies are used, as in standard 
angle-resolved photoelectron spectroscopy (ARPES) experiments 
\cite{Xu2013,Zhu2013,Neupane2013,Jiang2013,Denlinger2014,Xu2014}.

At the same time, a HAXPES spectrum of SmB$_6$ cannot be interpreted as representing directly 
the Sm $4f$ spectral weight since the photo-ionization cross-section of the Sm $4f$ states is not the 
only one which contributes to the spectrum. Other states, like the B $2s$ or Sm $5d$, $6s$ 
may also have comparable photo-ionization cross-sections when hard x-rays are used
\cite{Trzhaskovskaya2001,Trzhaskovskaya2002,Trzhaskovskaya2006}. Table I lists the 
photo-ionization cross-sections of the B $2s$, $2p$ and Sm $4f$, $5p$, $5d$, $6s$ orbitals as 
extracted or interpolated from the data \cite{Trzhaskovskaya2001, Trzhaskovskaya2002,Trzhaskovskaya2006}provided by Trzhaskovskaya \textit{et al.}.

\begin{table}
\caption{Subshell photo-ionization cross-section ($\sigma$) at 6.5 keV extrapolated from Ref. \onlinecite{Trzhaskovskaya2001,Trzhaskovskaya2002, Trzhaskovskaya2006}. 
$\sigma$ is divided by the number of electrons in the subshell. $\beta$ denotes the dipole parameter 
of the angular distribution. The cross-section for horizontal and vertical geometries are obtained by 
$\sigma$[1+$\beta$\{1/4+3/4cos(2$\theta$)\}]. Here $\theta$ is the angle between the photo-electron 
momentum and the polarization vector $E$ of the light. In the horizontal and vertical geometries, 
$\theta$=0 and 90 deg., respectively.}

\begin{ruledtabular}
\begin{tabular}{ccccc}
Atomic& $\sigma$/$e^-$ & $\beta$ & Horizontal & Vertical\\
subshell &(kb) & &(kb) &(kb)\\
\hline
B 2$s$ &1.462E-3 &1.945 &4.304E-3 &4.037E-5\\
B 2$p_{1/2}$ &6.303E-6 &0.015 &6.395E-6 &6.258E-6\\
Sm 4$f_{5/2}$ &4.935E-3 &0.547 &7.635E-3 &3.586E-3\\
Sm 5$p_{1/2}$ &8.753E-2 &1.540 &0.222 & 2.011E-2\\
Sm 5$p_{1/3}$ &7.365E-2 &1.634 &0.194 & 1.349E-2\\
Sm 6$s$ &9.467E-3 &1.942 &2.785E-2 &2.737E-4\\
Sm 5$d_{3/2}$ &0.013 &1.043 &2.616E-2 &6.127E-3\\
\end{tabular}
\end{ruledtabular}\label{tabl1}
\end{table}

In order to extract the more relevant Sm $4f$ spectral weight from HAXPES, we can make use of the 
pronounced dependence of the spectra on the polarization of the light as given by the so-called 
$\beta$-asymmetry parameter of the photo-ionization cross-sections of the various atomic shells involved
\cite{Trzhaskovskaya2001, Trzhaskovskaya2002,Trzhaskovskaya2006}. They are also listed 
in Table I. In particular, it has been shown experimentally by Weinen \textit{et al.} 
\cite{Weinen2015}, that the $s$ contribution to the spectra can indeed be substantially reduced 
(albeit not completely suppressed due to side-scattering effects) if the direction of the collected 
outgoing photoelectrons is perpendicular to the electric field vector of the light.

To make use of this polarization dependence we measured the valence band spectra of the 
\textit{in-situ} cleaved SmB$_6$ crystal using the two photoelectron energy analyzers, one 
positioned in the horizontal geometry, and the other mounted in the vertical geometry 
(see section \ref{sec-exp}). The spectra obtained in this manner are displayed in 
Fig. 4(a) in red and blue, respectively. The spectra are normalized with respect 
to the peak height of the features positioned at 0.1 and 1.1 eV binding energy. These features 
are known to originate from the Sm $4f$ states. We can clearly observe that there is a very 
strong polarization dependence in a very wide energy region of the spectra, i.e., from 3 eV 
to 12 eV binding energy. The difference between the two spectra is displayed by the green curve
in Fig. 4(b) and has maxima at about 5 and 10 eV. 

In order to elucidate the origin of this strong polarization dependence, we have
listed in Table I the effective photo-ionization cross-sections for the two geometries and 
performed band structure calculations using the full-potential non-orthogonal local orbital 
code (FPLO) \cite{Koepernik1999} to extract the B $2s$, $2p$ and Sm $4f$, $5p$, $5d$, $6s$ partial 
density of states (PDOS). The local density approximation (LDA) including spin-orbit (SO) coupling 
was chosen. We considered a non-magnetic calculation with the Sm $4f^6$ configuration\cite{Kasinathan2015}, 
and obtained a total DOS which is quite similar to an earlier calculation for the 
same Sm configuration \cite{Antonov2002}.  
The PDOSes are multiplied by the Fermi function and convoluted with a 0.2 eV FWHM Gaussian 
broadening, and shown in Fig. 4(b). Here we have weighted the relevant PDOSes with the following 
factors: from Table I we calculate the photo-ionization cross-sections relative to that of the Sm $4f$, 
and list them in Table II for each geometry; subsequently, we take the difference of the numbers 
between the two geometries and use them as multiplication factors for the PDOSes.
 
\begin{table}
\caption{Subshell photo-ionization cross-sections relative to that of Sm 4$f$. The horizontal and 
vertical cross sections in Table I are divided by those of Sm $4f$. The difference values are 
obtained by subtracting the numbers of the vertical from the horizontal.}

\begin{ruledtabular}
\begin{tabular}{cccc}
Atomic& Horizontal & Vertical &difference\\
subshell & & & \\
\hline
B 2$s$ &0.564 &1.126E-2 &0.552\\
B 2$p$ &8.376E-4 &1.745E-3 &-9.077E-4\\
Sm 5$p$ &2.727E+1 &4.686 & 2.258E+1\\
Sm 6$s$ &3.648 &7.633E-2 &3.571\\
Sm 5$d$ &3.426 &1.709 &1.718\\
\end{tabular}
\end{ruledtabular}\label{tabl2}
\end{table}

Fig. 4(b) compares the experimental horizontal-vs-vertical difference spectrum 
(green line) with the weighted PDOSes. The sum of these weighted PDOSes (black dashed line)
is in reasonable agreement with the experiment: the two main maxima at 5 and 10 eV energy
are reproduced. The fact that the intensity ratio between these two main maxima does not match 
well can perhaps be explained by the expected differences in the atomic orbitals used in the 
photo-ionization cross-section calculations compared to the ones used in the FPLO band structure code. 
We should note that we have artificially shifted the results of our calculations by 1 eV 
towards higher binding energies in order to better align the positions of the main features. This shift 
may be viewed as an \textit{ad-hoc} correction to the band structure calculations which did not take 
into account the intermediate valent state of Sm. 
It is also interesting to note that the photo-ionization cross-section numbers in Table I and II 
are extremely large for the Sm $5p$ in comparison to those of the other orbitals. Consequently, the inclusion 
of the Sm $5p$ becomes important for a quantitative analysis of the valence band HAXPES spectra, although
in terms of electronic structure, the contribution of the Sm $5p$ PDOS to the valence band can be
safely neglected.

Although the experimental valence band spectrum taken with the vertical geometry as
shown in Fig. 4(a) (blue line) represents already mainly the Sm $4f$ spectral weight (see Tables I and II), 
we nevertheless can make a further attempt to remove as much as possible the non-$4f$ contribution 
by carrying out the following exercise: we subtract from the vertical spectrum (I$_v$, blue line, Fig. 4a) the 
horizontal-vs-vertical difference spectrum (I$_h$-I$_v$, green line, Fig. 4b) multiplied by factor A, 
and we also subtract from the horizontal spectrum (I$_h$, red line, Fig. 4a) the same horizontal-vs-vertical 
difference spectrum (I$_h$-I$_v$, green line, Fig. 4b) but now multiplied by factor B, 
such that the so-obtained spectra are identical:  I$_v$ - A(I$_h$-I$_v$) = I$_h$ - B(I$_h$-I$_v$), i.e. B-A =1.
We have found A=0.8 and B=1.8 and we refer to the result as the extracted 4$f$ spectral weight represented by the
black line in Fig. 4(c). If the orbitals that made up the horizontal-vs-vertical difference spectrum were to have 
the same $\beta$ asymmetry parameter, then this procedure will remove the non-$4f$ contributions from the 
vertical and horizontal spectra. Fig. 4(c) displays this extracted $4f$ result (black line), together with the assignments 
of the atomic multiplet structure (blue sticks and labels) belonging to the photoemission final states which are 
reached when starting from the Sm$^{2+}$ and Sm$^ {3+}$ ground states \cite{Denlinger2014}. 
We can clearly see that the extracted 4$f$ spectral weight spectrum contains most of the sharp 
multiplet features, not only the high intensity ones at 0.1 and 1.1 eV but also smaller ones in the energy 
range between 3 and 12 eV. Obviously, there are also some 'left-over' intensities that do not match the 
multiplet structure. As explained above, the subtraction procedure cannot be perfect since the 
different non-$4f$ orbitals have different $\beta$ asymmetry parameters (see Table I). 

An important result to take from Fig. 4(a) and (c) is that there are only two main peaks in the energy 
range up to 2 eV, namely at 0.1 and 1.1 eV. This is to be contrasted to several photoemission studies 
using ultra-violet light where the presence of yet another peak at 0.8 eV binding energy has been 
reported \cite{Zhu2013,Neupane2013,Denlinger2014,Allen1980}. Based on our HAXPES results, 
we infer that this 0.8 eV peak very likely originates from the surface region of the SmB$_6$ material, 
supporting the assignment made earlier by Allen \textit{et al.} \cite{Allen1980}. In fact, the extreme 
sensitivity of this feature to the experimental conditions \cite{Zhu2013,Denlinger2014}, e.g. the rapid 
disappearance with time even under ultra-high vacuum conditions, suggests strongly that the 0.8 eV 
peak is caused by Sm atoms residing on top of the surface. Given the fact that the (001) surface 
investigated in the ARPES studies is polar \cite{Zhu2013}, a Sm termination must indeed be 
accompanied by a substantial electrostatic potential rearrangement for the Sm atoms at the
surface. Yet, STM studies also revealed that an unreconstructed Sm-terminated surface is rather rare. 
Instead, complex ordered and disordered surface structures are more commonly observed
\cite{Roessler2014,Roessler2016}.

\subsection{Surface of \textit{ex-situ} polished SmB$_6$}

\begin{figure*}[t!]
\includegraphics[scale=0.6]{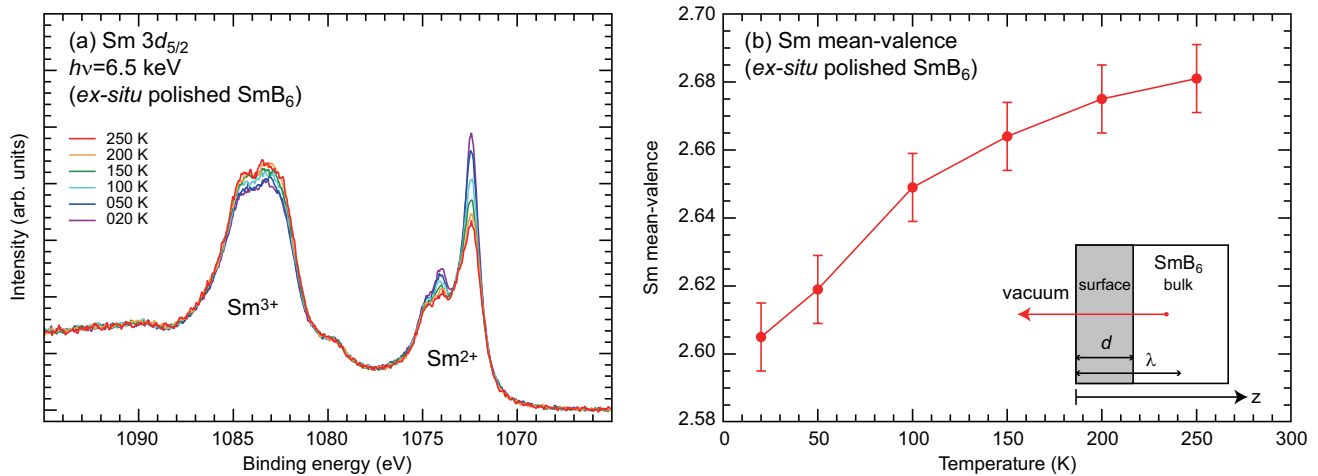}
\caption{(Color online) (a) Temperature dependence of the Sm 3$d_{5/2}$ spectra of the \textit{ex-situ} 
polished SmB$_6$ single crystal. (b) The temperature dependence of the Sm mean-valence as extracted 
from the Sm spectra in (a). The inset shows a schematic model of the bulk and surface regions as used 
in the fits, see text. Here, $d$ is the thickness of the surface region, $\lambda$ the inelastic mean-free 
path of the photoelectrons and $z$ the distance from the sample surface.} \label{fig.5}
\end{figure*}

One can readily expect that the surface of an \textit{ex-situ} polished SmB$_6$ single crystal will be 
different from the one of an \textit{in-situ} cleaved sample. Not only will any Sm present on the surface 
be oxidized, but also the oxidation process may in principle continue further into the bulk material, 
thereby creating a thicker surface region in which the Sm may have a valence different from the bulk 
value. In order to investigate the consequences of an \textit{ex-situ} preparation of the samples, 
we also carried out Sm $3d$ core-level photoemission studies on \textit{ex-situ} polished SmB$_6$ samples.

In Fig. 5(a) the Sm $3d_{5/2}$ spectrum of an \textit{ex-situ} polished SmB$_6$ and its temperature 
dependence is presented. It exhibits the same Sm$^{2+}$ and Sm$^{3+}$ components with the same 
temperature tendency as the \textit{in-situ} cleaved SmB$_6$. However, the Sm mean-valences are 
shifted to higher values over the entire temperature range as compared to those of the \textit{in-situ} 
cleaved sample: For the \textit{ex-situ} polished SmB$_6$ we obtained $v =$ 2.61 at 20 K and $v =$ 2.68 at 250 K, see Fig. 5(b). This is to be compared to 2.55 and 2.64, respectively, for the \textit{in-situ} cleaved SmB$_6$.

Clearly, an analysis of the spectra obtained for the \textit{ex-situ} polished SmB$_6$ sample has now 
to take into account the possibility of a non-uniform value of $v$ at the surface and in the bulk. 
To accomplish this, we adopt a minimal model (even simpler than the one used in Ref.\ \onlinecite{Lutz2016}) 
in which we assume that the sample can be divided into two regions, namely the surface region which 
has the Sm in its fully oxidized 3+ state, $v_{\rm surf} =$ 3, and the bulk region which has its pristine 
intermediate-valence properties $v_{\rm bulk}$, see the inset of Fig. 5(b). This allows us to set up an 
equation for the measured average valence $v_{\rm av}$ of the \textit{ex-situ} polished SmB$_6$ 
taking also into account the probing depth of the photoemission measurement:

\begin{equation}
v_{\rm av}\! \int_{0}^{\infty}\! e^{-z/\lambda}dz = v_{\rm surf}\! \int_{0}^{d}\! e^{-z/\lambda}dz + v_{\rm bulk}\! \int_{d}^{\infty}\! e^{-z/\lambda}dz.
\end{equation}

Here, $\lambda$ is the inelastic mean-free path of the photoelectrons, $z$ the distance from the surface, 
and $d$ the thickness of the surface region. After integration, one can obtain $d$ from

\begin{equation}
\frac{d}{\lambda} = \ln \left[ \frac{v_{\rm surf} - v_{\rm bulk}}{v_{\rm surf}-v_{\rm av}} \right].
\end{equation}

Using the experimental values of $v_{\rm bulk}$ (Fig. 3)\cite{note} and $v_{\rm av}$ (Fig. 5(b))\cite{note} as well as an 
estimated inelastic mean-free path of about $\sim$ 72 {\AA} for 5.5 keV photoelectrons \cite{Tanuma1994}, 
we arrive at a thickness $d \approx$ 9.5 {\AA} using the 20 K data and $d \approx$ 9.7 {\AA} at 250 K. Although the employed model is highly schematic and should not 
be taken literally, it provides the surprising indication that the thickness of the oxidized or 
chemically damaged surface region of the \textit{ex-situ} polished SmB$_6$ is 
rather small, about 1 nm. It appears that SmB$_6$ has a surface which is relatively 'leak-tight' 
against exposure to ambient atmosphere. One then might conjecture that this could explain
why many of the conductivity measurements carried out under ambient conditions exhibit a surprisingly high reproducibility \cite{Allen1979,Gorshunov1999,Flachbart2001,Hatnean2013, Zhang2013,
Kim2013,Wolgast2013,Kim2014,Wolgast2015,Thomas2016,Nakajima2016}. \\

\section{Summary}
We have performed bulk sensitive hard x-ray photoelectron spectroscopy measurements on \textit{in-situ} 
cleaved SmB$_6$ to elucidate the Sm valence and the Sm $4f$ spectral weight of the bulk material. 
The multiplet structure in the Sm $3d$ core level spectra provides a reliable base for an analysis of the 
valence. This analysis results in a value of $v =$ 2.55 at $\sim$ 5 K, which is close to the theoretical 
estimate for the border separating topologically trivial from topologically non-trivial SmB$_6$. 
The strong increase of the valence with temperature suggests that this is driven by the entropic gain 
in free energy due to the higher degeneracy of the magnetic Sm$^{3+}$ $4f^5$ state compared to 
the non-magnetic $4f^6$ singlet state of the Sm$^{2+}$. At elevated temperatures we 
can clearly observe in our spectra the presence of the Boltzmann occupation of the $J=1$ state of 
the Sm $4f^6$ configuration. The strong polarization dependence in the valence band 
spectra allowed us to extract the Sm $4f$ spectral weight, thereby disentangling surface from bulk 
contributions to the valence band spectra collected by ARPES. The measurements on \textit{ex-situ} 
polished SmB$_6$ single crystals revealed an oxidized or chemically damaged 
surface region which is surprisingly thin, of order 1 nm only.

\acknowledgments
We would like to thank Thomas Mende and Christoph Becker for their skillful technical assistance and 
Sahana R\"o\ss ler for helpful discussions. D. K. acknowledges funding from the Deutsche 
Forschungsgemeinschaft through SPP 1666. K-T. K. acknowledges support from the Max Planck-POSTECH 
Center for Complex Phase Materials and the NRF (Grant. No. 2016K1A4A4A01922028) funded by MSIP of Korea.

\appendix
\section{Reproducibility of the Sm 3$d$ spectra}
\begin{figure}[h!t!]
\includegraphics[scale=0.6]{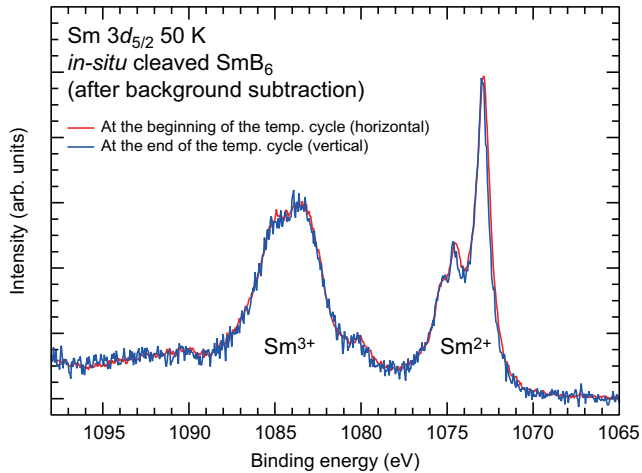}
\caption{(Color online) Sm 3$d_{5/2}$ spectra measured at 50 K at the beginning of the 
experiment (red line) and the end of the temperature cycle (blue line).}
\label{fig.A1}
\end{figure}

In order to verify the absence of surface degradation effects, we compare in Fig. \ref{fig.A1} the Sm 3$d_{5/2}$ spectrum measured 
at 50 K at the beginning of the experiment with the one measured at the very end of the temperature cycle 
(50 K$\rightarrow$20 K$\rightarrow$5 K$\rightarrow$100 K$\rightarrow$200 K$\rightarrow$250 K$\rightarrow$300 K$\rightarrow$50 K). 
The two spectra reproduce each other, thus demonstrating that surface degradation did not take place and that the observed
temperature evolution of Sm 3$d$ spectrum is real. The total measurement time for the cycle was 33 hours.

\section{Crystal electric field effect on the Sm$^{3+}$ 3$d$ spectrum}
In the case of Sm $f^5$ (Sm$^{3+}$), the lowest 4$f$ multiplet states are given by the $J$=5/2 and $J$=7/2,
with the latter about 130~meV higher in energy. A cubic crystal electric field splits the $J$=5/2 further into the 
quartet $\Gamma_8$ and the doublet $\Gamma_7$ states. 
Although the precise value of the crystal field for SmB$_6$ is still not known, if we adopt the value of the crystal field for NbB$_6$ determined from inelastic neutron scattering experiments\cite{Loewenhaupt1986}, the energy difference between the $\Gamma_8$ and $\Gamma_7$ states is about 13~meV, which is about one tenth of that between the $J$=5/2 and $J$=7/2 levels.

Assuming the same crystal field, we have calculated
the Sm $3d$ core-level spectrum for $T$=1~K and $T$=300~K. The results are shown in Fig. 7. In contrast with the Sm$^{2+}$ spectra, where we found the large temperature effects, we here clearly observe that the spectra are practically identical. One reason is that the energy splitting between $J$=5/2 and $J$=7/2 is too 
large to cause an appreciable Boltzmann occupation of the $J$=7/2 for the temperatures considered here, i.e. only
the $J$=5/2 contribute to the spectrum. Another reason is that the inclusion of the cubic crystal electric field does 
not add any noticeable new spectral features due to the fact that the $\Gamma_8$ and $\Gamma_7$ states originate from the 
same $J$ quantum number, while at the same time the crystal field energy scale is about two orders of magnitude 
smaller than that of the inverse life time of the $3d$ core-hole.

\begin{figure}[h!]
\includegraphics[scale=0.6]{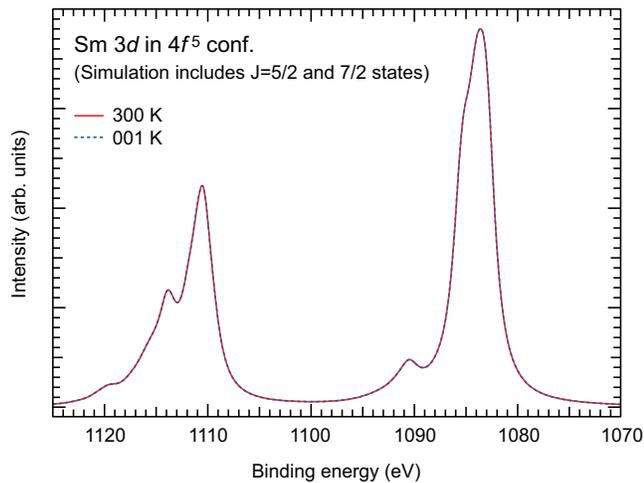}
\caption{(Color online) Calculated Sm 3$d$ core-level spectrum for a Sm$^{3+}$ ion
at $T$=1~K (blue dashed line) and $T$=300~K (red line) with the cubic crystal field for NdB$_6$ in Ref.~\onlinecite{Loewenhaupt1986}. The $J$=5/2 $\Gamma_8$ ground state, the $J$=5/2 $\Gamma_7$ excited state at $\sim$13~meV and higher lying $J$=7/2 excited states around 130~meV are included for the initial state (see text). The spectra were convoluted with a Lorentzian function with 
FWHM=0.45 eV and a Gaussian function with FWHM=0.22 eV.}
\label{fig.A3}
\end{figure}

\section{Background correction for the Sm 3$d$ spectra}
\begin{figure*}[t!]
\includegraphics[scale=0.6]{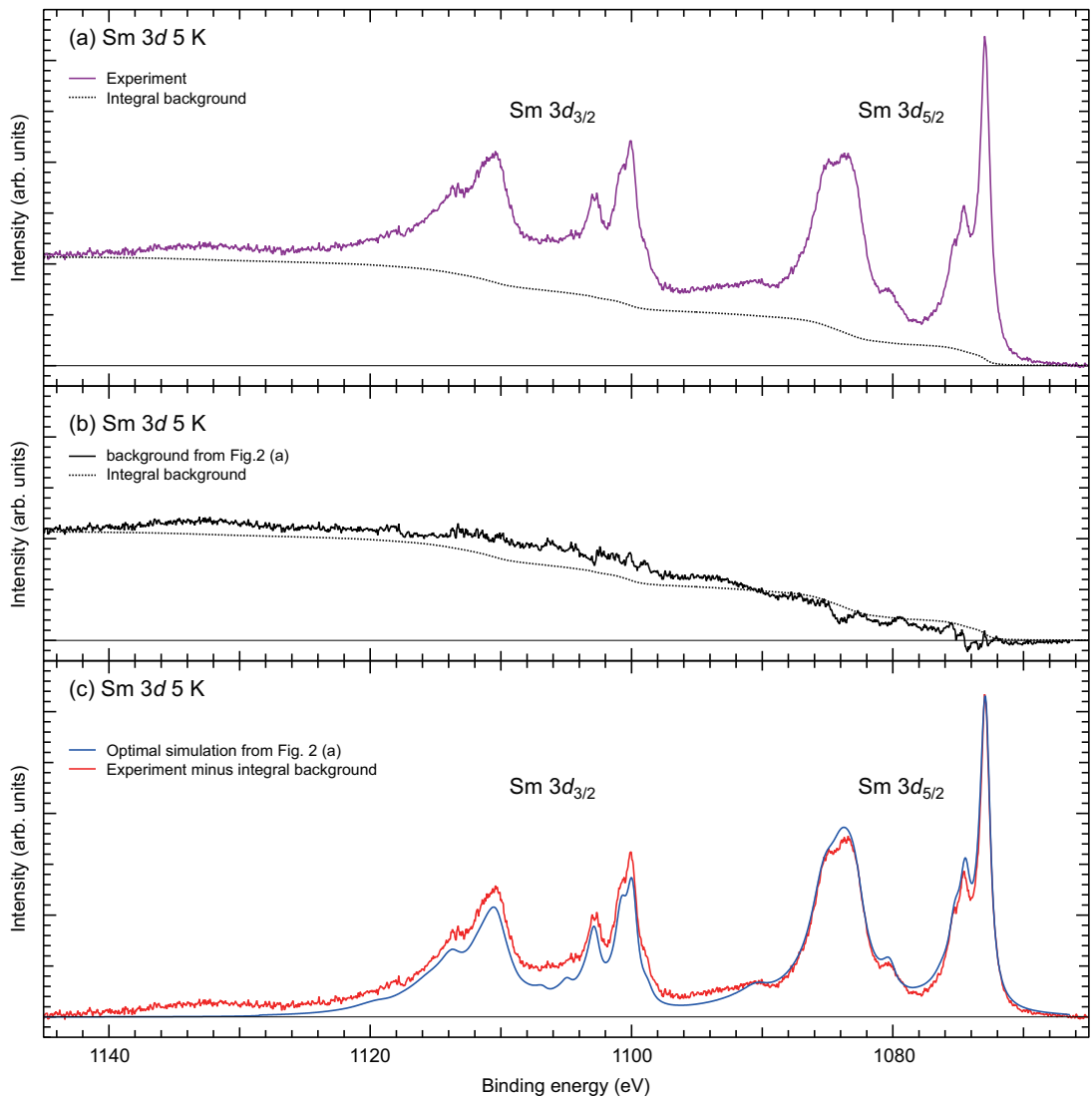}
\caption{(Color online) (a) The experimental spectrum (purple line) of the in-situ 
cleaved sample taken at 5 K together with its integral-background (black dashed line).
(b) Comparison of the integral background (black dashed line) with the background 
(black line) from Fig. 2 (a). 
(c) Experimental spectrum corrected for the integral background (red line) and the 
optimal simulation (blue line) from Fig. 2 (a).}
\label{fig.A2}
\end{figure*}

 The standard procedure in the literature in evaluating the valence of mixed valent 
strongly correlated systems from core level spectra is to first make a correction for the 
background signal due to inelastic electron scattering processes, and then to evaluate 
the intensities of the relevant configurations, in our case, the Sm$^{2+}$ and Sm$^{3+}$.
The problem is that for this procedure to work accurately one needs to know the loss-function 
(in photoemission) in order to know what line shape the background should have. However, 
the loss-function is usually not known and it is a major effort to determine it experimentally. 
It is obvious that different assumptions for the line shape of the background will lead to different 
background-corrected spectra and thus likely to different values for the valence.
To illustrate the ambiguities that enter when using a background correction procedure, 
we now apply the generally used integral background correction \cite{Shirley1972} 
to our 5 K spectrum, see panel (a) of Fig. \ref{fig.A2}. It is interesting to note that
this integral background shows discrepancies to the background that we have obtained 
using the multiplet line shape analysis as displayed in Fig. 2 (a), see panel (b) of Fig. \ref{fig.A2} 
and compare the black dashed line with the black line, respectively. Consequently, there 
are also discrepancies between the integral-background-corrected spectrum and the optimal 
simulation from Fig. 2 (a), i.e. compare the red line with the blue line, respectively, in panel (c) of 
Fig. \ref{fig.A2}. The integral-background-corrected spectrum has in fact intensities over a wide 
energy range that cannot be accounted for by the multiplet structures. Also the intensity of the 
3$d_{3/2}$ relative to the 3$d_{5/2}$ has increased in the integral-background corrected spectrum 
in comparison with the multiplet theory, meaning that the integral-background corrected 
spectrum violates the atomic branching ratio between 3$d_{5/2}$ and 3$d_{3/2}$ components. 
This indicates that our multiplet line shape analysis can give a more reliable Sm 
valence value than the one using integral background.

\begin{thebibliography}{99}
\bibitem{note} For the estimation of the thickness of the surface region, we used 
$v_{\rm bulk}$=2.549 and 2.635 at 20 K and 250 K, and $v_{\rm av}$=2.605 and 2.681 at 20 K and 250 K, 
respectively. These are the same values as listed in the text, but are now presented with one digit more as
they come out from the fitting analysis. Although the error bar for the absolute value of the estimated Sm 
valence is 0.02, the ratio of the valences may benefit from a more accurate representation since systematic
errors in the fitting analysis may cancel out.
\end{thebibliography}
\end{document}